\documentclass[12pt]{iopart}

\usepackage{iopams,graphicx}
\usepackage{color}  
\begin{document}

\title[]{Metallo-dielectric core-shell nanospheres as building blocks for optical 3D isotropic negative-index metamaterials}

\author{R. Paniagua-Dom\'inguez$^1$, F. L\'opez-Tejeira$^1$, R. Marqu\'es$^2$ and J. A. S\'anchez-Gil$^1$}

\address{$^1$ Instituto de Estructura de la Materia,
Consejo Superior de Investigaciones Cient{\'\i}ficas, Serrano 121,
28006 Madrid, Spain}
\address{$^2$ Departamento de Electr\'onica y Electromagnetismo, Universidad de Sevilla, 41012, Sevilla, Spain}
\ead{j.sanchez@csic.es}
\begin{abstract}
Materials showing electromagnetic properties that are not attainable in naturally occurring media, the so called 
metamaterials, have been lately, and still are, among the most active fields in optical and materials physics and engineering. 
Among those properties, one of the most attractive is the sub-diffraction resolving 
capability predicted for media having index of refraction of -1. Here we propose a fully 3D, isotropic metamaterial 
with strong electric and magnetic responses in the optical regime, based on spherical metallo-dielectric core-shell nanospheres. 
The magnetic response stems from the lowest, magnetic-dipole resonance of the dielectric shell with high refractive index,
and can be tuned to coincide with the plasmon resonance of the metal core, responsible for the electric response.  
Since the response does not originate from coupling between structures, no particular periodic arrangement needs to be imposed. Moreover,
due to the geometry of the constituents, the metamaterial is intrinsically isotropic and polarization independent.
It could be realized with current fabrication techniques with materials such as Silver (core) and Silicon or Germanium (shell). For these
particular realistic designs, the metamaterials present negative index in the range within $1.2-1.55$~$\mu$m. 

\end{abstract}

\maketitle

\section{Introduction}
Once the possibility of 
building a negative-index metamaterial (NIM) was proven in the microwave regime \cite{Smith}, extraordinary effort 
has been made to obtain analogous behaviours for increasingly higher frequencies up to the 
visible range of the electromagnetic spectrum \cite{ShalaevNatPhot,SouNatPhot}. 
The major obstacle found when trying to translate
ideas to higher frequencies was how to achieve a strong diamagnetic response in the systems designed.
Many attempts to tackle the problem were simple
miniaturizations of the canonical designs employed in those first metamaterials operating in microwaves. As an example,
the split-ring configurations, or slight variations of it, were reduced up to the nanoscale to obtain such a magnetic response. Apart from the fundamental 
limitations inherited from those designs, e.g. anisotropy, and the increasingly complexity in the fabrication procedures, several drawbacks have been found in 
this process, some of them being consequence of the different behaviour of metals when excited with higher 
frequency waves \cite{PRLSoukoulis}.
Precisely this different behaviour also inspired some authors to search for 
different configurations intended to exploit the plasmonic response to obtain artificial 
magnetism. In many of them artificial magnetism is due to coupling between different plasmonic structures, their 
drawback thus being the high losses within metallic parts \cite{Engheta}. In most cases, the proposed designs are restricted to operate under
certain polarization and incidence conditions \cite{PRBSoukoulis}-\cite{AtwaterOE}, or are not truly three-dimensional materials \cite{AtwaterNatMat,AtwaterSci}. 
Moreover, in many cases, the behaviour of the proposed designs stems from coupling between the different constituents, thus making particular
arrangements necessary. As a consequence, spatial dispersion effects often appear due to propagation of waves in the lattice.
Lately, some approaches based on plasmonic waveguides supporting
negative-index modes have pushed the frequencies in which left-handed (LH) behaviour is obtained well within the visible spectrum. 
However, propagation inside such material would be limited to the propagation length of the plasmon in the waveguide,
thus making the design a single layer device, lacking truly three-dimensionality \cite{AtwaterNatMat,AtwaterSci}. In addition, various attempts 
have been made towards NIM by exploiting magnetic resonances occurring in structures made of high permittivity materials \cite{Wheeler}-\cite{Vynck}. Some
of them combine these structures with secondary structures providing the electrical response \cite{Yannopapas05}-\cite{Kussow} or embed them
in a metallic host medium \cite{SheoItoh}, thus having inherent high losses.    
Here we report a design that tackle many of the previously mentioned limitations. A totally three-dimensional isotropic negative-index metamaterial
 operating at optical frequencies, and whose response is due to every isolated
``meta-atom``. Therefore, no particular arrangement of them is needed. We study the possiblity
of using a spherical core-shell 
configuration to obtain with one single structure both electric and magnetic responses. The core, 
being metallic, is responsible for the electrical response, while the shell, made of a high 
permittivity dielectric, provides the strong diamagnetic response. An extension of Mie 
theory is exploited to rigorously determine the scattering properties and resonances of the whole spherical core-shell configuration,
which essentially determine the effective material properties \cite{Lederer}.
Calculations for core-shell structures built up with realistic materials (Ag@Si and Ag@Ge) demonstrate the possibility to obtain NIM operating 
at 1.2 $\mu$m-1.55$\mu$m. Since both responses are attained directly from one single constituent, no particular arrangement of the inclusions is 
needed. In order to extract the effective parameters of the metamaterial, we will assume both a random and a simple cubic distribution.
In the case of a random distribution Lorentz-Lorenz theory is applied, 
leading to simultaneously negative effective permeability and permittivity for several filling fractions. 
In the case of a simple cubic lattice arrangement, finite-element-method is applied to carry out numerical simulations, which 
fully account for interaction between the periodically arranged constituents. 
Effective material constants are then extracted through standard S-parameter \cite{RetPar,RobRetPar}
retrieval procedure, and tested to fulfill causality and passivity, thus confirming a true double-negative index.
Due to the spherical symmetry of the constituents, the metamaterial response will be essentially isotropic and polarization independent.

\section{Optical properties of metallo-dielectric core-shell meta-atoms}
Let us examine the  scattering of a plane electromagnetic wave (wavelength $\lambda$) from a spherical core-shell 
particle without any approximation, which indeed can be done analytically as an 
extension of Mie theory, obtained first by Aden and Kerker \cite{AK}. Figure \ref{Fig1}(a) depicts the geometry of one of the basic constituents or ``meta-atoms'': a high permittivity ($\epsilon$)
 dielectric shell is considered with outer radius $R\ll\lambda$ and thickness $T$, $\epsilon_c$ and $\epsilon_0$
being the dielectric constant of the core and the surrounding medium, respectively.
The scattering and extinction efficiencies can be expressed in terms of the material and geometrical parameters through the
scattering coefficients $a_l$ and $b_l$ (which represent, respectively, the different electric and magnetic multipolar contributions) as:
\begin{eqnarray}\label{eq_Mie} 
Q_{sca}=\frac{2}{y^2}\sum_{l=1}^{\infty}(2l+1)(|a_l|^2+|b_l|^2)\\
Q_{ext}=\frac{2}{y^2}\sum_{l=1}^{\infty}(2l+1)\Re(a_l+b_l),
\end{eqnarray}
where $y=kR$. The mentioned scattering coefficients can be written in terms of the 
the spherical Bessel functions of the first ($j_l(x)$) and second ($y_l(x)$) class and depend on $\epsilon_c/\epsilon_0$, $\epsilon/\epsilon_0$, 
$R_{in}$ and $R$. $\Re$ denotes the real part. Their explicit form is:
\begin{equation}
a_l=\frac{\psi_l(y)[\psi'_l(ny)-A_l\chi'_l(ny)]-n\psi'_l(y)[\psi_l(ny)-A_l\chi_l(ny)]}{\xi_l(y)[\psi'_l(ny)-A_l\chi'_l(ny)]-n\xi'_l(y)[\psi_l(ny)-A_l\chi_l(ny)]}
\end{equation}
\begin{equation}
b_l=\frac{n\psi_l(y)[\psi'_l(ny)-B_l\chi'_l(ny)]-\psi'_l(y)[\psi_l(ny)-B_l\chi_l(ny)]}{n\xi_l(y)[\psi'_l(ny)-B_l\chi'_l(ny)]-\xi'_l(y)[\psi_l(ny)-B_l\chi_l(ny)]}.
\end{equation}
The Ricatti-Bessel functions introduced are $\psi_l(z)=z j_l(z)$, $\chi_l(z)=-z y_l(z)$ and
$\xi_l(z)=z h_l^{(1)}(z)$, where $h_l^{(1)}(z)=j_l(z)+i y_l(z)$ is the spherical Hankel function of the first class. The coefficients $A_l$ and $B_l$ are:
\begin{equation}
A_l=\frac{n\psi_l(nx)\psi'_l(n_cx)-m_c\psi'_l(ny)\psi_l(n_cx)}{n\chi_l(nx)\psi'_l(nx)-n_c\chi'_l(ny)\psi_l(n_cx)} 
\end{equation}
\begin{equation}
B_l=\frac{n\psi_l(n_cx)\psi'_l(nx)-n_c\psi_l(ny)\psi'_l(n_cx)}{n\chi'_l(nx)\psi_l(n_cx)-n_c\psi'_l(n_cy)\chi_l(nx)},
\end{equation}
where $x=kR_{in}$, $n^2=\epsilon/\epsilon_0$ and $n_c^2=\epsilon_c/\epsilon_0$. It is inmediate to realize that 
all information about material and geometrical properties of the core is contained in these $A_l$ and $B_l$ coefficients.

\subsection{Magnetic resonance of a high permittivity shell}
\begin{figure}[t]
\centering
\includegraphics[width=1.0\columnwidth]{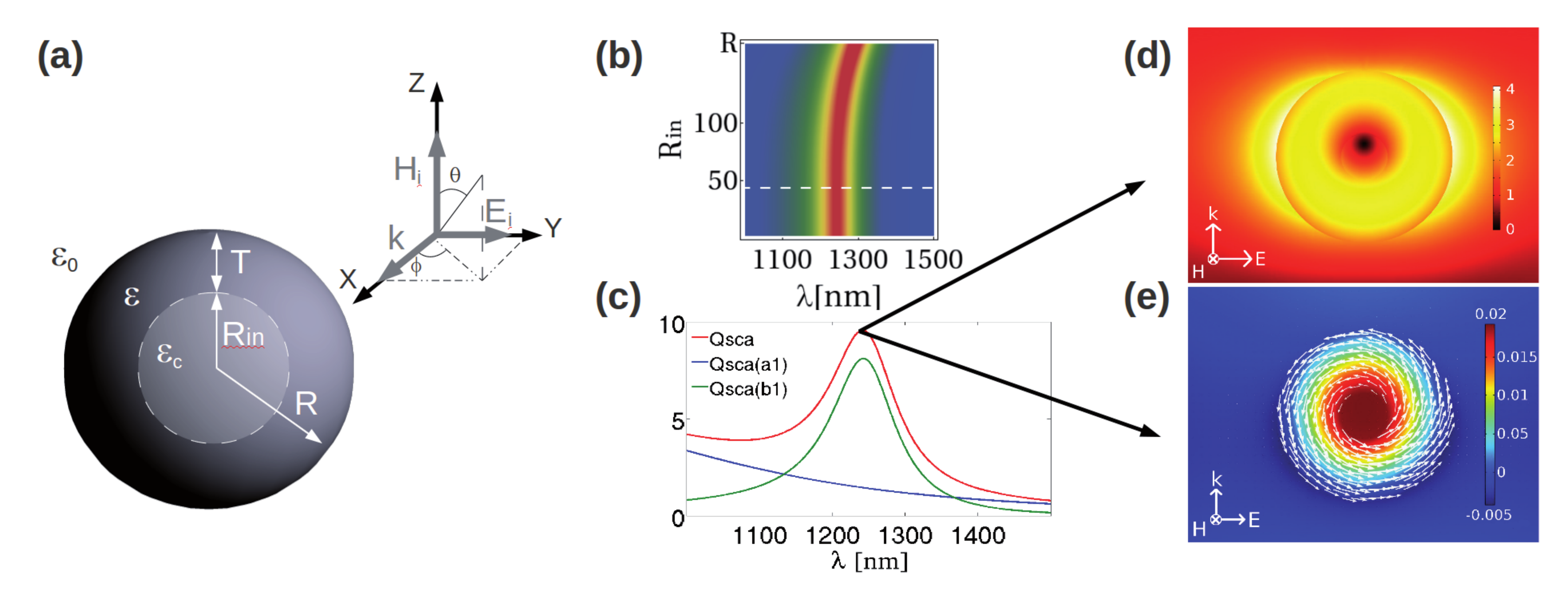}
\caption{(a) The geometry of the problem. (b) Dipolar magnetic
contribution to the total scattering efficiency of a Silicon (Si) nanoshell with outer radius $R=170$~nm as a function of the inner radius ($R_{in}$) and wavelength of the incident light. (c) Total scattering efficiency,
together with the dipolar electric ($a_1$) and magnetic ($b_1$) contributions to the scattering efficiency for inner
radius $R_{in}=45$~nm (indicated in (b) by a dashed white line). (d-e) Near-field plots 
at the magnetic resonance for the Si shell of (c). (d) Norm of electric 
field. (e) Out-of-plane component (only non-zero component of the incident magnetic field) of 
$\mathbf{H}$, together with the electric displacement field in white arrows.}
\label{Fig1}
\end{figure}

To better understand the physics behind, we first characterize the magnetic resonance of a nanoshell for a real dielectric material with 
high  refractive index. We choose Silicon, Si, the refractive index of which can be considered 
constant, $n=\sqrt{\epsilon/\epsilon_0}\sim~3.5$, and lossless, within the 
near-IR range $\lambda=1\mu m-2\mu m$.
In figure \ref{Fig1}(b) the contribution is plotted to the total scattering efficiency of the magnetic dipolar term ($b_1$) 
as a function of $R_{in}$ and the incidence wavelength. It is the dominant contribution,
as can be seen in figure \ref{Fig1}(c) for the specific case of $R_{in}=45$~nm.
A resonance can be clearly seen, the wavelength of which corresponds to that of the Si compact sphere \cite{OESaenz} 
when $R_{in}\rightarrow 0$ and starts to redshift when the shell thickness is thin enough ($T\lesssim R/3$) and decreases. 
Thus the magnetic resonance of the Si nanoshell can be tuned within a certain range of 
wavelengths larger than that of the Si sphere resonance.

The behaviour of the EM fields at the magnetic resonance has been calculated by
full-wave (finite-element method) numerical simulations \cite{Comsol}. The result is 
depicted in figure \ref{Fig1}(d) and (e). It can be observed that the electric displacement field rotates in 
planes parallel to the equator, thus inducing a strong magnetic moment.
This pattern clearly shows that the electrical displacement current $-i\omega\varepsilon{\bf E}$ is strongly confined inside the shell, and
rotates along the $\phi$ direction around the incident H-field. Such behaviour reveals that the resonance is qualitatively quite similar to the
first resonance of a dielectric ring, already reported \cite{MarquesJCondMatt}, following an LC model with an inductance $L\propto\mu_0 R$ associated to
the circulation of the displacement current, and a capacitance $C\propto\varepsilon R$ associated to the electric energy confined in the shell. At
resonance, a strong magnetic moment along the direction of the incident H-field is generated in the same qualitative way as in the dielectric ring analyzed \cite{MarquesJCondMatt}. 
\subsection{Electric and magnetic resonances on core-shell systems}
Next we analyze the mutual influence of a metallic core and a dielectric shell. 
With regard to the plasmon resonance of the metallic core (i.e., collective
oscillations of conduction electrons), it is well 
known that for small spheres ($R_{in}\ll \lambda$) of dielectric permittivity 
$\epsilon_c(\omega)$, embedded in a medium with dielectric constant $\epsilon_0$, 
the induced dipole moment is resonant at the frequency $\omega_{LSPR}$ such that  
$\epsilon_c(\omega_{LSPR})=-2\epsilon_0$ \cite{BH}. 
Therefore, when the metallic sphere is coated with a thick dielectric layer of permittivity 
$\epsilon$, the  quantity driving the resonance condition is not $\epsilon_c$ itself but, 
instead, the ratio between the permittivities of the core and the coating. 
That is, for a small metal core, the resonance occurs when $\epsilon_c/\epsilon\sim-2$ is 
fulfilled, thus redshifting the localized surface-plasmon resonance (LSPR) with respect to that 
in vacuum. 

The question now arises as to whether or not the shell magnetic resonance is preserved when the core is 
metallic. Since the confinement and rotation of the electric field inside the coating is directly 
related to the jump conditions for the normal component of the field between the shell and the 
surrounding medium \cite{MarquesJCondMatt}, and since the metal is expected to avoid penetration of the field inside, no 
significant changes of the magnetic resonance behaviour are expected at least for relatively small metallic cores. This will be revealed
below, where we plot the scattering efficiency together with the electric 
and magnetic dipolar contributions.
Such behaviour makes possible to still predict the appearance of the magnetic resonance even 
if a metallic core is present. Furthermore, the electric and magnetic responses of a metal-dielectric core-shell 
nanosphere can be tuned as to make both resonances coincide. Let us take silver (Ag) as the 
material to build up the core and Si again as a high permittivity dielectric for the shell. Complex Ag permittivity values are taken from
tabulated data \cite{JC}. Our goal is to obtain simultaneously both resonances in the near infrared (IR) range of the 
electromagnetic spectrum.  The LSPR  will be redshifted to that wavelength such that 
$\epsilon_r^{Ag}\sim-2\epsilon^{Si}$. This happens at a wavelength about 720~nm. 
Nevertheless, we have to keep in mind that, strictly, this would only be a good approximation for 
very small metallic particles with relatively thick coatings. Since we want to push the electric 
resonance to the IR to make it coincide with a magnetic one stemming from the Si shell,
we can take increasingly larger radius for the core. When the size of a metallic particle is increased, 
there is a redshift in the resonance wavelength that can be explained in terms of depolarization 
effects \cite{BH}.
\begin{figure}[t]
\centering\includegraphics[width=1.0\columnwidth]{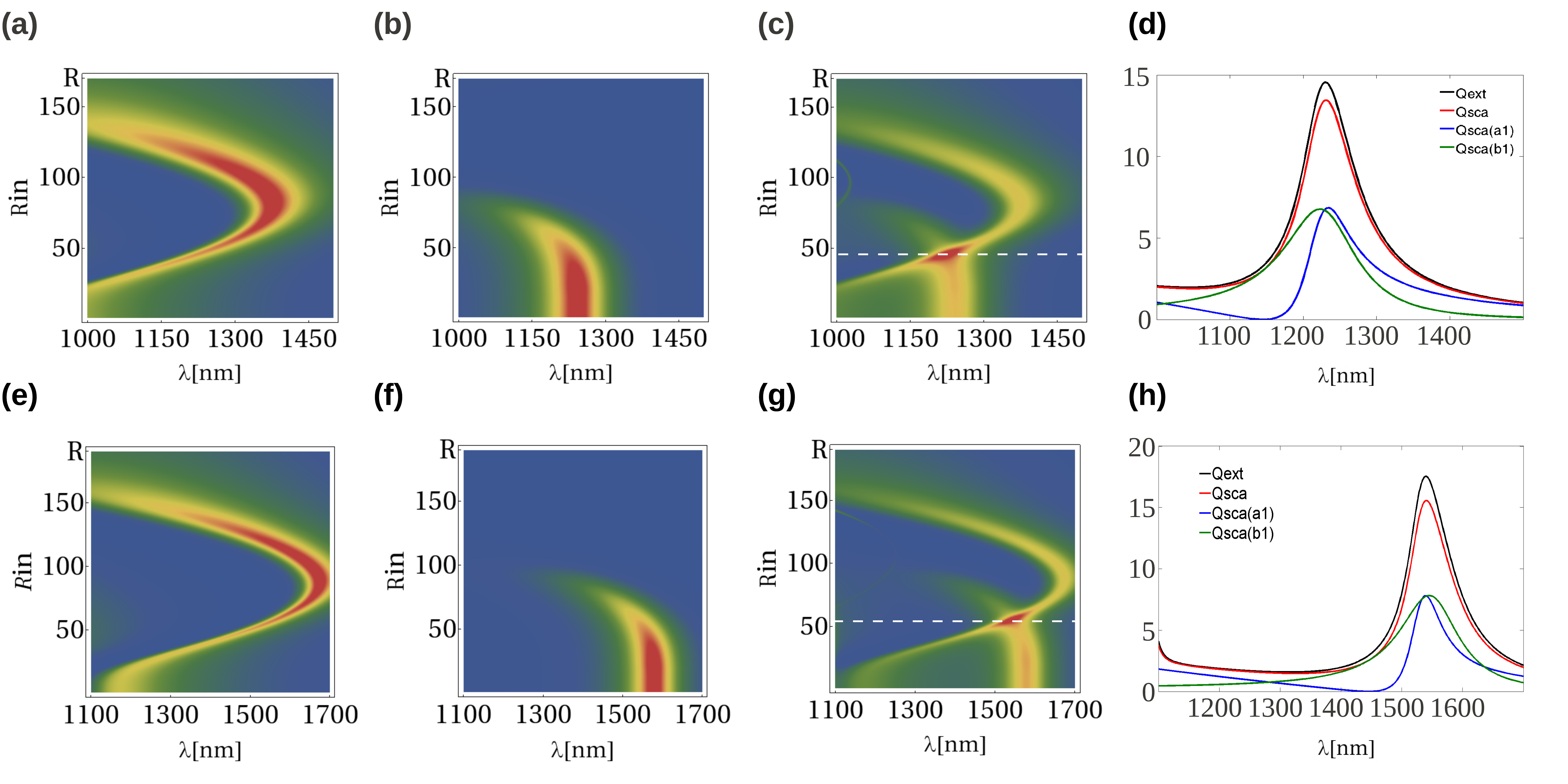} 
\caption{Optical properties of Ag@Si and Ag@Ge core-shell nanospheres. (a-c) Dipolar electric contribution (a), dipolar magnetic contribution (b),
 and total scattering efficiency (c) of a Ag@Si core-shell nanosphere with outer radius $R=170$~nm as a function of the inner 
radius ($R_{in}$) and wavelength of the incident light. (d) Extinction and scattering efficiencies, together with the dipolar electric ($a_1$)
and magnetic ($b_1$) contributions to the scattering efficiency for inner radius $R_{in}=47$~nm (indicated in (c) by a dashed white line).
(e-g) Dipolar electric contribution (e), dipolar magnetic contribution (f),  and total scattering efficiency (g) of a Ag@Ge core-shell 
nanosphere with outer radius $R=190$~nm as a function of the inner radius ($R_{in}$) and wavelength of the incident light. (h) Extinction and 
scattering efficiencies, together with the dipolar electric ($a_1$) and magnetic ($b_1$) contributions to the scattering efficiency for inner 
radius $R_{in}=55$~nm (indicated in (g) by a dashed white line).}
\label{Fig2}
\end{figure}

Figures \ref{Fig2}(a)-(c) depicts both, the electric dipolar and the magnetic dipolar contributions to the 
scattering efficiency, together with the total scattering efficiency for a Ag@Si  
core-shell system of outer radius $R=170$~nm as a function of $R_{in}$ and the incidence wavelength. It can be clearly seen
an overlap between the electric and magnetic resonances. It happens within 1150~nm and 1300~nm and for inner radius between $R_{in}=40-50$~nm. 
We have explicitly plotted the case $R_{in}=47$~nm (figure \ref{Fig2}(d)). 
An interesting feature that can be observed is that
the magnetic resonance disappears as the inner radius increases. It can be explained by the fact that the electric displacement field in the core 
rotates in the opposite direction as it does in the shell, thus reducing the total magnetic moment generated.
Another effect can be seen in the electric dipole contribution. As the inner radius increases the resonance broadens and redshifts as
expected. Interestingly, this behaviour changes when the thickness is comparable with $R_{in}$, and the resonance starts to blueshift. This
effect can be attributed to the fact that, as the dielectric shell becomes thinner, the resonance wavelength tends to that of a sphere without
coating, thus blueshifting. Finally, it is important to note that the specific wavelengths at which the electric and magnetic resonances overlap
are determined by the geometrical parameters of the structure, as well as by the specific materials used. Therefore, it is possible
to tune the wavelength of overlapping by appropriate choice of these parameters. This gives the design a great degree of freedom, and make it
plausible to operate at different frequencies. As an example, a Ag@Ge system with $R=190$~nm is presented (figures \ref{Fig2}(e)-(h)). In this
case resonances overlap within 1500~nm and 1620~nm for $R_{in}=50-60$~nm. The particluar case with $R_{in}=55$~nm is plotted in figure \ref{Fig2}(h).

\begin{figure}[t]
\centering\includegraphics[width=1.0\columnwidth]{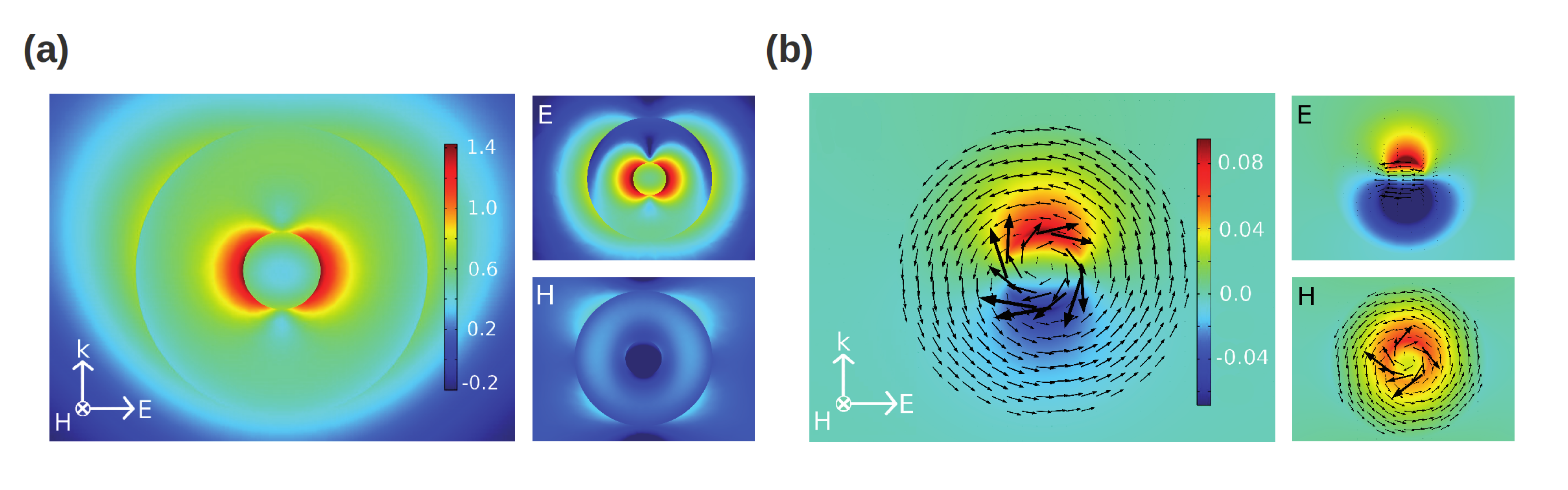} 
\caption{Near-field plots for a Ag@Si shell with $R=170$ nm and $T=123$ nm. (a) Norm of the electric field in $Log_{10}$ scale at combined electromagnetic resonance and
when the system is excited by a purely electric excitation (up-right, E) and by a purely magnetic one (down-right, H). Colour scale, incidence and polarization are preserved in all figures. (b) Out-of-plane component of $\mathbf{H}$, together with 
the electric displacement ($\mathbf{D}$) field in black arrows at combined electromagnetic resonance and when the system is excited by a purely electric excitation (up-right, E) or a purely magnetic one (down-right, H). Colour scale, incidence and polarization are preserved in all figures.}
\label{Fig3}
\end{figure}

In figure \ref{Fig3}, the near-field pattern at the combined magnetic (shell) and electric (core) 
resonance is shown for the Ag@Si system. Also plotted are the responses of the system to a purely electric and
 magnetic excitation. These were obtained by placing a perfect mirror 
at a distance $3\lambda/4$ and $\lambda$, respectively, from the center of the structure. The
 distinctive behaviour of both contributions is preserved in the combined electromagnetic resonance, revealed
 through the rotating field confined within the Si shell (as in figure \ref{Fig1}(e)), together with the dipolar LSPR resonance of the Ag core. Note that,
indeed, the electric displacement in the core and the shell rotates in opposite directions and that the magnetic
 near field pattern, figure \ref{Fig3}(b), can be explained as a combination of the electric and magnetic contributions. Near-field patterns for 
core-shell with Ge covers are similar to those shown here.

\section{Calculation of effective parameters}

What do we expect for a material consisting of such core-shell nanostructures?
In general it is highly non-trivial to extract the effective constitutive parameters of 
a metamaterial. Here, two different methods are applied to obtain the effective parameters of
a metamaterial composed by the core-shell structures presented. In the first one we will assume
a random arrangement of the constituents, while in the second a metamaterial made of a cubic arrangement
of them will be studied, both leading to a negative-index behaviour within certain wavelengths.

\subsection{Effective parameters of a random arrangement of core-shell nanospheres}

\begin{figure}
\centering
\includegraphics[width=0.9\columnwidth]{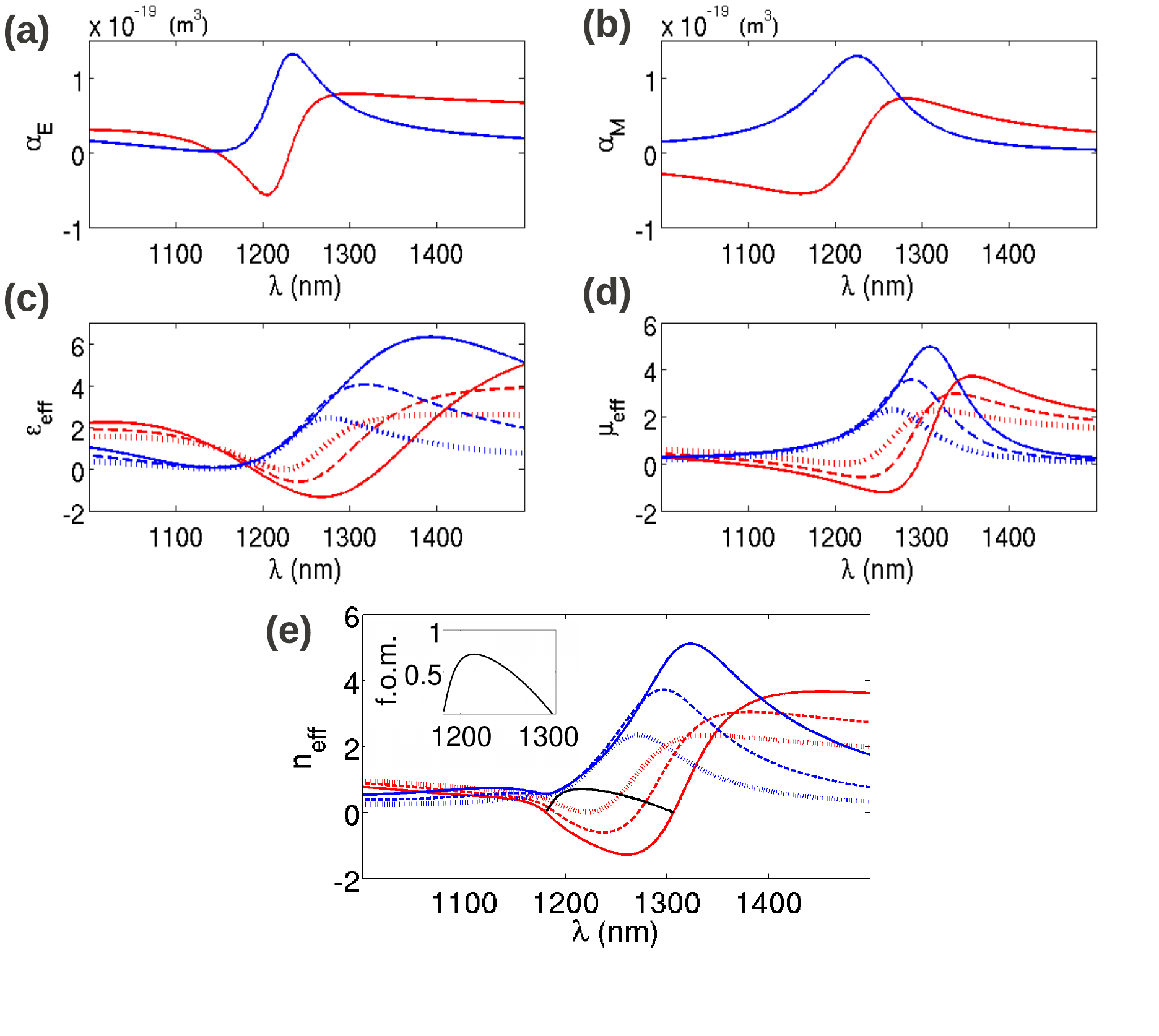}
\caption{(a)-(e). Electric and magnetic polarizabilities for a Ag@Si core-shell with $R_{in}=47$~nm and $R=170$~nm and effective parameters for
a metamaterial composed of a random arrangement of these structures. (a) Real (red) and imaginary (blue) parts of the electric polarizability.
(b) Real (red) and imaginary (blue) parts of the magnetic polarizability. (c)-(e) Real (red) and imaginary (blue) parts of the effective permittivity (c), permeability (d) and refractive index (e) of a metamaterial with 
several filling fractions: $f=1/3$ (dotted), $f=0.5$ (dashed) and $f=2/3$ (continuous). The 
black curve and the inset in (e) represent the $f.o.m.$ for the highest filling fraction in the left-handed spectral region.} 
\label{Fig4}
\end{figure}

\begin{figure}
\centering
\includegraphics[width=0.9\columnwidth]{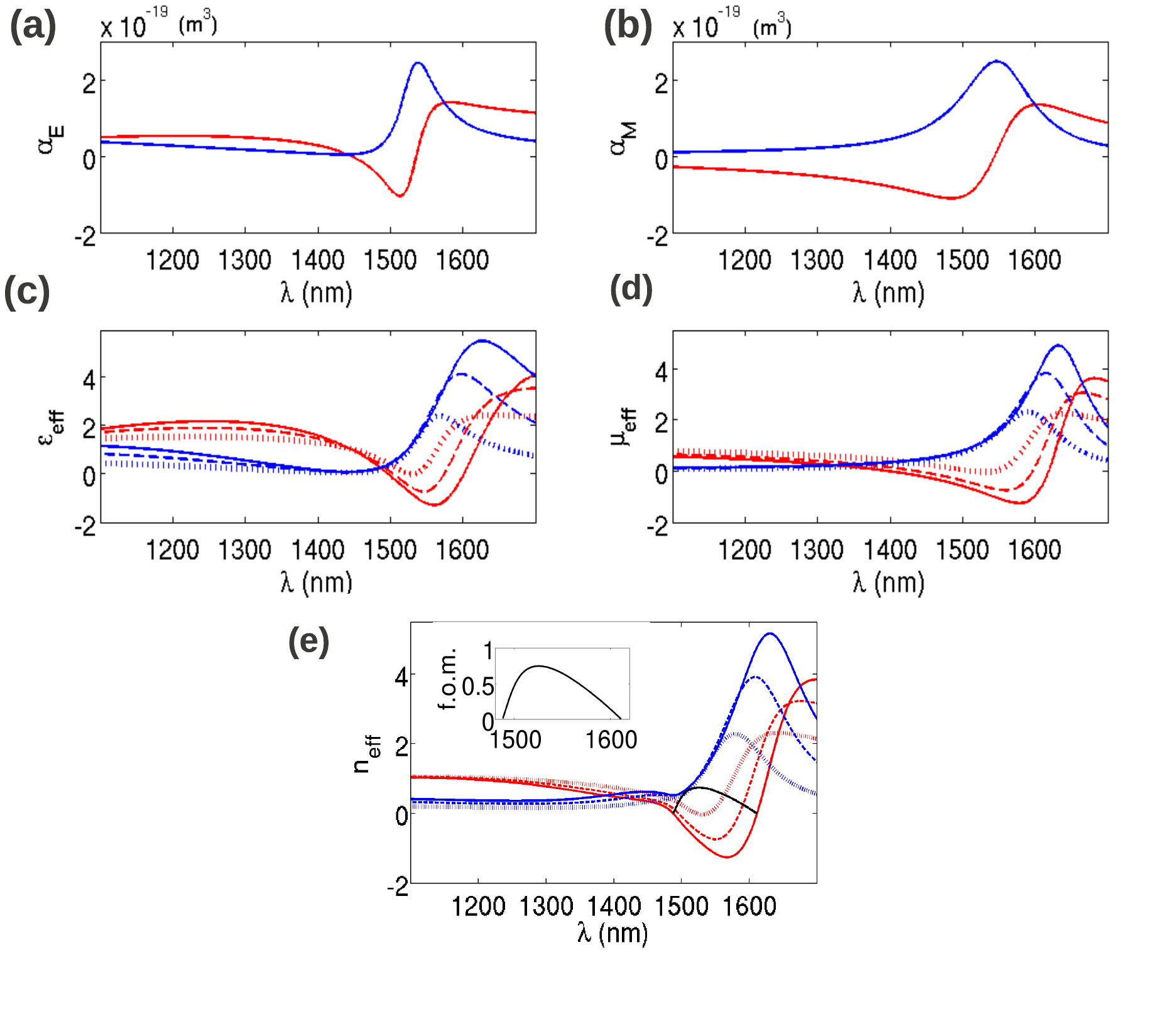}
\caption{(a)-(e) Electric and magnetic polarizabilities for a Ag@Ge core-shell with $R_{in}=55$~nm and $R=190$~nm and effective 
parameters for a metamaterial composed of a random arrangement of these structures. (a) Real (red) and imaginary (blue) parts of the electric 
polarizability. (b) Real (red) and imaginary (blue) parts of the magnetic polarizability. (c)-(e) Real (red) and imaginary (blue) parts of 
the effective permittivity (c), permeability (d) and refractive index (e) of a metamaterial with several filling fractions: $f=0.25$ (dotted), $f=0.4$ (dashed) and $f=0.5$ (continuous). 
The black curve and the inset in (e) represent the $f.o.m.$ for the highest filling fraction in the left-handed apectral region.} 
\label{Fig5}
\end{figure}

For composites made of a cubic or random arrangement of dipolar particles, Lorentz-Lorenz theory is widely used 
\cite{Wheeler,Wheeler2}, leading to the well known Clausius-Mossotti 
formulas relating the effective permittivity and permeability with the polarizabilities
of the particles and the filling fraction $f=(4/3)\pi NR^3$, where $N$ is the number of particles per unit volume:
\begin{equation}\label{CM}
 \frac{\epsilon_{\mathtt{eff}}-\epsilon_0}{\epsilon_{\mathtt{eff}}+2\epsilon_0}=f\frac{\alpha_E}{4\pi R^3},\ \ \ \ \ \ \frac{\mu_{\mathtt{eff}}-\mu_0}{\mu_{\mathtt{eff}}+2\mu_0}=f\frac{\alpha_M}{4\pi R^3}
\end{equation}
where $\alpha_{E}$ and $\alpha_{M}$ are the electric and magnetic polarizabilities of the spherical particles.
For the core-shell structures considered $R/\lambda\sim1/7$. Therefore, we expect them to be well within the approximation considered in the theory, 
thus behaving essentially as electric and magnetic point dipoles. 
The electric and magnetic polarizabilities, $\alpha_E$ and $\alpha_M$, respectively,  
are directly proportional to the scattering coefficients $a_1$ and $b_1$ [factor $i(k^3/6\pi)^{-1}$].
In figure \ref{Fig4}(a) and (b), we have plotted the polarizabilities for a core-shell configuration of $R_{in}=47$~nm and $R=170$~nm. 
It is clear from the graph that there is a spectral region where negative electric and magnetic polarizabilities are obtained.
Interestingly, as mentioned before, no particular arrangement for the constituents is necessary to build the 
metamaterial, since the resonant electric and magnetic responses 
arise from each core-shell structure separately. Thus, assuming a random distribution
we can compute the effective parameters from (\ref{CM}) for different filling fractions. Figures \ref{Fig4}(c)-(e) depict the calculated
effective permittivity, permeability, and refractive index for metamaterials made up by these Ag@Si core-shells with filling 
fractions $f=1/3$, $f=0.5$ and $f=2/3$. The metamaterial has simultaneously negative permittivity and permeability for filling
 fractions higher than $f=1/3$. For a filling fraction of $f=2/3$, 
the system, moreover, has $|n_{\mathtt{eff}}|\sim 1$, although this relatively high filling fraction would be
in the limit of validity of Clausius-Mossotti formulas. In order to quantify the losses of the system one can compute the so called 
figure of merit, defined as $f.o.m.=|\Re(n_{eff})|/\Im(n_{neff})$. 
The computed $f.o.m$  in the left-handed spectral region for this filling fraction
is depicted by the black curve, reaching a maximum value of $f.o.m.\sim 0.71$, corresponding
to $\Re (n_{eff})\sim -0.8$. If Ge is used instead of Si, the NIM behaviour starts with lower filling fractions, 
due to a stronger electric and magnetic responses (see figure \ref{Fig5} and compare the polarizabilities with those of figure \ref{Fig4}). Superlensing capabilities 
are predicted for filling fractions lower than $f=0.5$ (figures \ref{Fig5}(c)-(e)). Again, the $f.o.m.$ for the highest filling fraction
is plotted as a black curve. In this case it reaches a maximum value of $f.o.m.\sim 0.75$, corresponding to $\Re (n_{eff})\sim -0.5$. Although being
far from the best $f.o.m.$  values reported for double-fishnet metamaterials ($f.o.m.\sim 3)$ \cite{WegenerOL,FishnetGM}, the predicted values are reasonably
good, with the obvious advantage of isotropy of this proposal. 

\subsection{Effective parameters of a cubic periodic arrangement of core-shell nanospheres}

To further test the metamaterial design, we consider now core-shell nanospheres arranged periodically
 in the vertices of a cubic lattice. We take Germanium as the high permittivity coating and
choose the geometrical parameters as $R=190$~nm and $R_{in}=55$~nm. We consider
the case in which the cubic lattice has a period $d=385$~nm (corresponding to a
filling fraction $f\sim0.5$). Full numerical simulations (see section Methods) of propagation across infinite slabs 
of different thicknesses ranging from 9 to 13 unit cells are carried out to ensure convergence. Bloch boundary conditions are applied in the interfaces with adjacent cells. 
Therefore, coupling between structures (with an interparticle distance of only 5~nm) is fully considered. The spectral range of the simulation is 
1.28 to 1.7~$\mu$m and the wave impinges at
normal incidence. The effective material parameters are retrieved from the complex reflection and transmission coefficients through 
the usual equations found in the literature \cite{RetPar,RobRetPar}. 
\begin{figure}
\centering
\includegraphics[width=0.9\columnwidth]{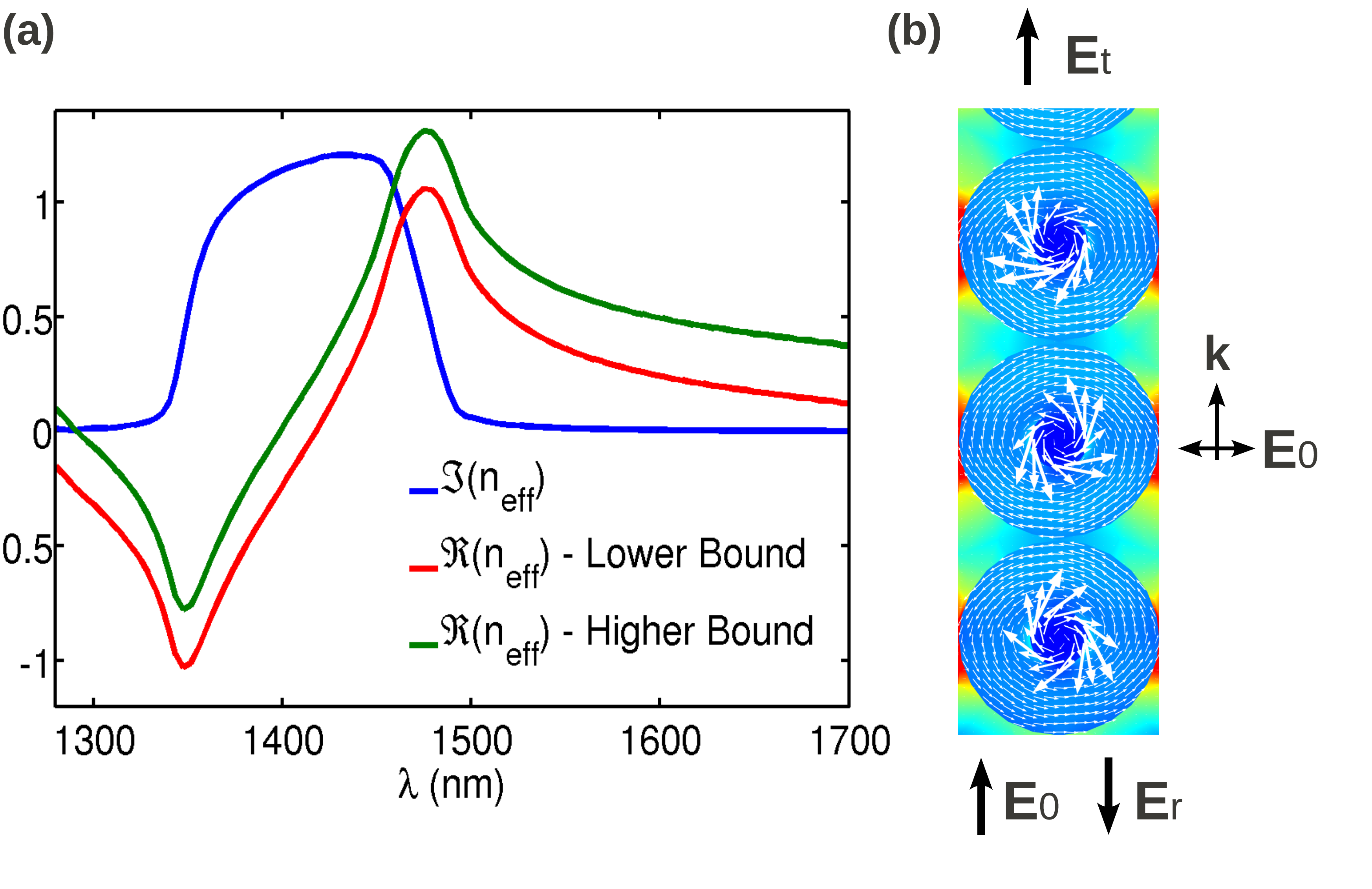}
\caption{(a) Real and imaginary parts of the retrieved effective index of refraction of a metamaterial 
made by Ag@Ge core-shell nanospheres of $R=190$~nm and $R_{in}=55$~nm arranged
in the vertices of a cubic lattice with period $d=385$~nm, corresponding to a filling fraction $f\sim0.5$. (b) Detail
of the normalized near electric field norm (color scale, red$\rightarrow$1, blue$\rightarrow$0) and electric displacement 
field (indicated by white arrows) in the periodic structure.} 
\label{Fig6}
\end{figure}
While the imaginary part is unambiguously computed from the simulation, once a sufficiently high number
of unit cells in the propagation direction have been considered, the real part of the index of refraction must be determined
with much more care. Concretely, one must pay special attention to ensure that the retrieved refractive index fulfills causality
and passivity. In this way we have obtained two bounding values for the real part of the index, so it fulfills the two mentioned
requirements. For a detailed description on how the index was obtained, see section Methods.
In figure \ref{Fig6}, the obtained bounds for the refractive index are plotted. With this retrieval technique, a negative index spectral region
is predicted. In the more conservative prediction, corresponding to the higher bound, the real part of the index
reaches a minimum value of $n_{\mathtt{eff}}\sim-0.77$,
with a $f.o.m.\sim1.88$, corresponding to a wavelength of 1.347~$\mu$m. Although the values for the index are lower than those predicted by Clausius-Mossotti,
the negative index frequency band is present and confirms the core-shell design as meta-atom candidate for building a
negative index metamaterial. It needs to be pointed out that we are far from the limit of filling fractions that can be obtained with non-overlapping
spheres (cannonball pile). Since for interparticle distances as low as 5~nm, we found that coupling does not prevent
the appearance of the magnetic resonance, we expect to reach higher negative values of the index in other periodic configurations with
higher filling fractions. Further study of periodic configurations with other homogenization approaches is intended for future work.

\section{Concluding remarks and fabrication possibilities}

Obtaining a three-dimensional isotropic metamaterial having negative index of refraction in the optical part of the spectrum has been
one of the major challenges for scientist and engineers devoted to electromagnetism in the last decade. Here we presented a
new design based on core-shell nanospheres that operates in the near-infrared, which tackle many of the previously found problems, namely 
isotropy, polarization Independence and lack of three-dimensionality.
In our system, the effective response of the metamaterial is due to every isolated ``meta-atom''. Therefore,
no particular arrangement of the constituents is needed. Specifically, we demonstrated with realistic materials that, for a random arrangement,
the system achieves double-negative index of refraction for different filling fractions and that super-resolution is possible. We also tested
the validity when a very simple periodic realization is assumed. Although the achieved values in the latter case are worse than predicted by 
Clausis-Mossotti for the same filling fractions,
we haven't explored here some other periodic configurations that are expected to give a stronger response. 

With regard to building these metamaterials, current Silicon fabrication techniques allow the realization of complex
nanostructures such as hollow nanospheres \cite{nanoshells} and opals \cite{Opals}. In
some of the processes, the starting point are Silica (SiO$_{2}$) nanostructures, as is the case in \cite{Opals}, in which Si opals are
fabricated by magnesiothermic reduction of SiO$_{2}$ opals. Since silver nanospheres have been successfully covered with SiO$_{2}$ in variable thicknesses \cite{JACS},
there are plausible ways to realize the metamaterial proposed here, at least with Si covers. Concerning the fabrication of Ge shells instead, a layer
of a different material, suitable to grow it, can be added between the core and the shell. This layer, if thin, would not affect excessively the physical response of the system,
opening the possibility of fabrication of this system as well. The  underlying physical principles can of course be exploited at lower (far-IR and terahertz) frequencies, at which
some dielectric materials exhibit very large refractive indices
and certain materials (e.g. polar crystals, doped semiconductors) behave as plasmonic metals. 
Therefore, the results presented pave the way towards potential isotropic three-dimensional
optical metamaterials designed on the basis of the physics underlying the doubly-resonant 
metallo-dielectric configuration.

\section{Methods}

Extinction and scattering cross sections based on the extended Mie theory were calculated with Wolfram Mathematica 8. The near-field plots in figures
\ref{Fig1} and \ref{Fig3} were calculated using the RF module of COMSOL Multiphysics v4.0a. The 
computational domain consisted of four concentric spheres which defined four subdomains. The radii of the spheres were $R_{in}$, $R$,
$2R$ and $2.5R$. From inner to outer, the subdomains represented the core, the shell and 
the embedding medium (air). The last domain was set to a spherical PML, which absorbed all scattered radiation. The incident radiation was 
defined as a plane wave. The mesh was constructed with the software built-in algorithm, which generates a free mesh consisting on tetrahedral elements.
The maximum element edge size was set to 15~nm in the whole core-shell structure, with a growth rate of 1.35, meaning that elements adjacent 
to a given one should not be bigger than 1.35 times the size of it. For the PML and air domains, the maximum element edge size was
50~nm. All mesh sizes are below the value recommended in the program specifications, which sets a maximum edge size of  $1/10$ of the effective 
wavelength for a correct meshing. Finally, for the simulation of wave propagation along a N-unit-cell thick slab, COMSOL was also used.
In this case, the computational domain consisted
of two concentric spheres representing the core-shells, and a right rectangular prism of width and height $d=385$~nm, and length $L=Nd$. Bloch periodic boundary conditions were set in the
directions perpendicular to the propagation and two ports activated in the direction of propagation that allow to compute the reflection
and transmission complex coefficients. In these cases the maximum element edge size was set to 40~nm in the core-shell subdomains, 50~nm in the ports
boundaries, and 100~nm in the boundaries where periodic boundary conditions were applied. A correct solution of the problem requires an identical meshing
for the pairs of boundaries where periodic boundary conditions are applied. In all cases PARDISO solver was used. As an example, for 9-unit-cell slab
simulation, the total mesh consisted of 149064 elements, and the calculation involved 948768 degrees of freedom, requiring almost 18 GB of memory.
\subsection{Effective index retrieval procedure}
The effective refractive index for the metamaterial structure, shown in figure \ref{Fig6}, were calculated from complex transmission and reflection coefficients
following \cite{RobRetPar}. As mention before, we performed simulations of wave propagation across infinite slabs of different thicknesses. After a sufficient number
of unit cells is considered we get convergence of the results. We needed to consider 
up to 9 unit cells in the direction of propagation to get convergence. Then, we performed the simulation for 11 and 13 unit-cell-thick slabs. 

\begin{figure}
\centering
\includegraphics[width=1\columnwidth]{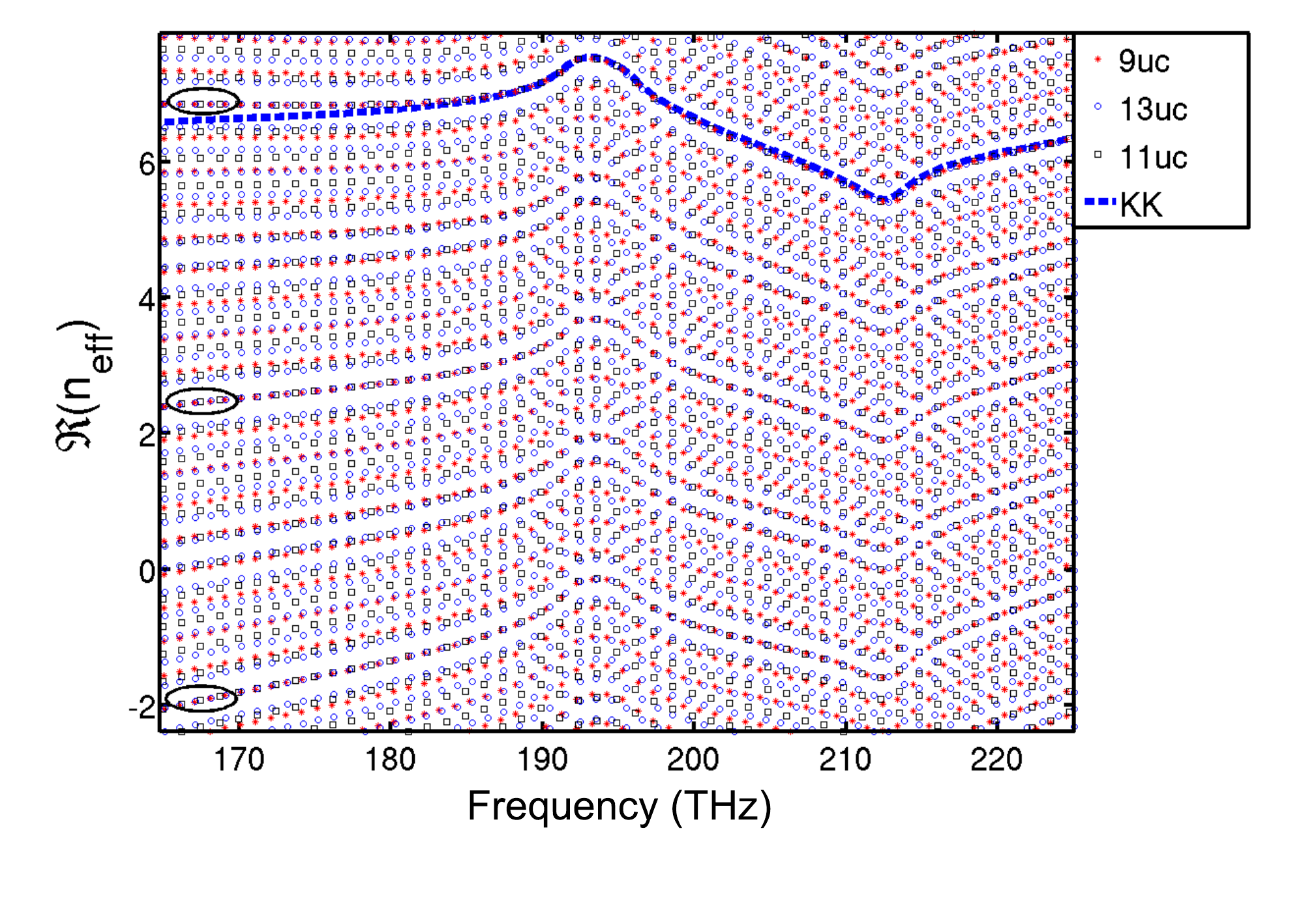}
\caption{A very large number of branches of the real part of the retrieved refractive index for 9 (red asterisks), 11 (black squares) and 13 (blue circles) unit-cell-thick metamaterial slabs, composed
by Ag@Ge core-shell nanospheres of $R=190$~nm and $R_{in}=55$~nm arranged
in the vertices of a cubic lattice with period $d=385$~nm, corresponding to a filling fraction $f\sim0.5$. The branches that are independent of the thickness
of the slab are indicated by a surrounding ellipse. The dashed blue line represents the real part obtained from the imaginary 
part by Kramers-Kronig relations, plus a constant factor to fit the only thickness-independent index showing good agreement.} 
\label{Fig7}
\end{figure}

It is well known that, when extracting the effective index through transmission and reflection coefficients, an ambiguity arises related to the branches of the complex
logarithm function. This ambiguity only affects the determination of the real part of the index, the imaginary one being univocally defined. It is a common assumption that
the real part of the index can be found plotting different branches for different thicknesses. The physical index is assumed to be the one independent of the 
thickness of the slab. However, this assumption may be ambiguous as well. The reason is that, if one plots a sufficiently high number of branches one will find
more than one unique index independent of the thickness. In figure \ref{Fig7} this situation is shown. We have plotted a very large number of branches corresponding to
 the Ag@Ge core-shell periodic configuration of section 3.2.  It can be clearly seen that there is not only one case in which the index is independent of the thickness.
Therefore, although this requirement of thickness-independence of the retrieved index is clearly physically acceptable, we should not take it as a
sufficient condition.

In fact, we should impose more conditions to the retrieved index. It must, no doubt, fulfill the basic requirement of causality, expressed mathematically by the Kramers-Kronig relations. Since, fortunately, we can unambiguously determine the imaginary part of the index, it is possible to apply Kramers-Kronig relations to get the real part of the index \cite{KK}. We need
to point out that, actually, by this technique one can only compute the corresponding real part up to a constant factor. The value of this factor, interpreted physically as 
the value of the index when the frequency tends to infinity, needs to be determined somehow else. The question is, then, if any of the indices that do not depend on thickness also fullfils causality. In figure \ref{Fig7}, the real part of the index, retrieved through Kramers-Kronig relations 
applied to the imaginary part, is plotted on top of the only thickness-independent index for which the fitting is in good agreement (the constant factor added is 6.35). 
It is apparent that the agreement is quite good. It should be noticed that this index is the only one for which the fitting is good, thus fulfilling the
the causality constrain.

However, one more constrain must be imposed in order to get an acceptable effective index. The resulting
 effective medium should be passive. This condition demands no spontaneous generation of energy, and mathematically relates the real and imaginary parts of the
 effective index, with the real and imaginary parts of the effective impedance of the slab. This impedance can be computed also
 from the complex reflection and transmission coefficients \cite{RobRetPar}. Applying this condition we find that none of the indices
retrieved directly from the transmission and reflection complex coefficients fulfill both causality and passivity.
Nevertheless, one can still compute a physically acceptable effective index of refraction. If one uses the passivity condition,
it is possible to impose upper and lower bounds to the constant factor in Kramers-Kronig relations. An index with
a real part given directly by Kramers-Kronig relations, applied to the univocally computed imaginary part, 
and a constant factor between the two bounds will necessarily fulfill both conditions. We found that, in fact, the upper and the 
lower bounds imposed by passivity are quite close in value, being a restrictive condition on the possible values of the index of refraction.
The two boundary values for the index are plotted in figure \ref{Fig6}(a).

\ack 
The authors acknowledge support both from the Spain Ministerio de Ciencia e Innovaci\'on through the
Consolider-Ingenio project EMET (CSD2008-00066) and NANOPLAS (FIS2009-11264), and
from the Comunidad de Madrid (grant MICROSERES P2009/TIC-1476). R.~Paniagua-Dom{\'{\i}}nguez acknowledges support from CSIC through a JAE-Pre grant.
We are indebted to an anonymous referee for helpful suggestions on the convergence of the COMSOL results in figure 6.

\end{document}